\documentclass{nature_pics}

\usepackage{graphicx}
\usepackage{amsmath}
\usepackage{wasysym}
\usepackage[utf8]{inputenc}
\usepackage[right,modulo]{lineno}

\title{Coherent diffractive imaging of single helium nanodroplets with a high harmonic generation source}

\author{Daniela Rupp$^{1}$, Nils Monserud$^2$, Bruno Langbehn$^1$, Mario Sauppe$^{1}$, Julian Zimmermann$^{1}$, Yevheniy Ovcharenko$^{1}$, Thomas M{\"o}ller$^{1}$, Fabio Frassetto$^{3}$, Luca Poletto$^{3}$, Andrea Trabattoni$^{4}$, Francesca Calegari$^{4,5}$, Mauro Nisoli$^{5,6}$, Katharina Sander$^{7}$, Christian Peltz$^{7}$, Marc J. J. Vrakking$^{2}$, Thomas Fennel$^{2,7}$ \& Arnaud Rouz\'{e}e$^2$}

\begin{document}
\maketitle

\begin{affiliations}
 \item Institut f{\"u}r Optik und Atomare Physik, Technische Universit\"at Berlin, Hardenbergstraße 36, 10623 Berlin, Germany
 \item Max-Born-Institut f{\"u}r Nichtlineare Optik und Kurzzeitspektroskopie, Max-Born-Straße 2A, 12489 Berlin, Germany
 \item CNR, Instituto di Fotonica e Nanotecnologie Padova, Via Trasea 7, 35131 Padova, Italy
 \item Center for Free-Electron Laser Science, DESY, Notkestr. 85, 22607 Hamburg, Germany
 \item CNR, Instituto di Fotonica e Nanotecnologie Milano, Piazza L. da Vinci 32, 20133 Milano, Italy
 \item Department of Physics, Politecnico di Milano, Piazza L. da Vinci 32, 20133 Milano, Italy
 \item Institut f{\"u}r Physik, Universit{\"a}t Rostock, Albert-Einstein-Straße 23, 18059 Rostock, Germany
\end{affiliations}

\begin{abstract}
Coherent diffractive imaging of individual free nanoparticles has opened novel routes for the in-situ analysis of their transient structural, optical, and electronic properties. So far, single-shot single-particle diffraction was assumed to be feasible only at extreme ultraviolet (XUV) and X-ray free-electron lasers, restricting this research field to large-scale facilities. Here we demonstrate single-shot imaging of isolated helium nanodroplets using XUV pulses from a femtosecond-laser driven high harmonic source. We obtain bright wide-angle scattering patterns, that allow us to uniquely identify hitherto unresolved prolate shapes of superfluid helium droplets. Our results mark the advent of single-shot gas-phase nanoscopy with lab-based short-wavelength pulses and pave the way to ultrafast coherent diffractive imaging with phase-controlled multicolor fields and attosecond pulses.
\end{abstract}

Single-shot coherent diffractive imaging (CDI) with intense short-wavelength pulses became possible just recently with the advent of XUV and X-ray free-electron lasers (FEL)~\cite{Chapman2010}. This lensless imaging method has revolutionized the structural characterization of nanoscale samples including biological specimens\,\cite{Seibert2011}, aerosols\,\cite{Loh2012}, atomic clusters\,\cite{ClustersAtFLASH, Gomez2014, Barke2015}, and nanocrystals~\cite{Xu2014}. By capturing high quality diffraction patterns from a single nanoparticle in free flight using a single laser pulse, the sample morphology can be determined in-situ and free from spurious interactions due to deposition on a substrate. For sufficiently regular structures the wide-angle scattering information even reveals the full three-dimensional particle shape and orientation~\cite{Barke2015,Xu2014, Reines2009}, as multiple projections of the same particle are encoded in a single diffraction image~\cite{Barke2015}. These unique capabilities enable the investigation of metastable or transient states that exist only in the gas phase. Pioneering FEL experiments have explored this frontier and demonstrated CDI of quantum vortices in helium droplets~\cite{Gomez2014}, ultrafast nanoplasma formation~\cite{Bostedt2012}, and explosion of laser-heated clusters~\cite{Gorkhover2016}. Using XUV and soft X-ray high harmonic generation (HHG) sources for single-shot nanoparticle CDI holds the promise to combine the nanoscale structural imaging capabilities of CDI with the exquisite temporal, spectral, and phase control inherent in the use of optical lasers, including the fascinating prospect of CDI with isolated attosecond pulses.

The brightness of HHG sources is typically orders of magnitude lower than that of an FEL~\cite{Miao2015}, but over the years a number of them have been scaled up in order to achieve high intensities and/or high average power~\cite{Hergott2002, Hadrich2011, Rudawski2013, Takahashi2013, Hong2014, Popmintchev2015}, allowing coherent diffractive imaging of fixed targets~\cite{Miao2003, Sandberg2007, Ravasio2009, Chen2009, Seaberg2011, Zuerch2014}. Experiments on ion-beam edged nanostructures in membranes achieved impressive resolution on the order of 20\,nm after multiple exposures~\cite{Seaberg2011, Zuerch2014} and even yielded the overall shape from a single-shot diffraction image in favorable cases~\cite{Ravasio2009}. Here we report the feasibility of lab-based 3D characterization of unsupported nanoparticles and demonstrate single-shot HHG-CDI on individual free helium nanodroplets.

In our experiment (Fig.\,1), a high power Ti:sapphire laser amplifier was used to generate 35\,fs laser pulses at 792\,nm wavelength with up to 33\,mJ pulse energy. About 12\,mJ were loosely focused ($f = 5$\,m) into a xenon-filled gas cell~\cite{Schuette2014}, producing $\approx 2$\,$\mu$J of XUV radiation, i.e. close to $10^{12}$ photons per pulse. This corresponds to approximately $1 \%$ of the pulse energy that can currently be achieved at the XUV free-electron laser FERMI\,\cite{Allaria2012, FermiUserWebpage}. The CDI application requires high fluence and thus tight focusing. Typical back-reflection multilayer mirrors, however, conflict with the use of straylight apertures and the detection of scattered light at small scattering angles. Therefore, CDI-compatible grazing-incidence microfocusing optics with an overall transmission of $10\%$ based on a coma-correcting system of toroidal mirrors\,\cite{Frassetto2014} were used to focus the multicolor XUV beam to a small spot (beam-waist $\omega_0=10$\,$\mu$m), achieving a power density of $3\times10^{12}$ W/cm$^2$ (pulse averaged). The XUV spectrum (11$^{\text{th}}$ to 17$^{\text{th}}$ harmonic, see Fig.\,3d) was obtained prior to the CDI measurements with a grating spectrometer. A jet of helium droplets with diameters of several hundreds of nanometers crossed the focus of the XUV beam. The droplets were generated using a cryogenically cooled pulsed valve maintained at a temperature between 4.9\,K and 5.7\,K, operating at low repetition rates of 3-10\,Hz. The diffracted radiation was measured shot-to-shot with a wide-area MCP (micro-channel plate) based scattering detector (see Methods).

Within $3\times10^5$ single-shot measurements, 2300 bright patterns with distinct structures were obtained and another 12700 recorded images contained weak, unstructured scattering signal. A selection of exemplary diffraction patterns is displayed in Fig.\,2. To analyze the diffraction patterns, the multicolor components of the HHG pulses have to be taken into account. The simultaneous use of several different wavelengths complicates the analysis of the size and shape of the particles, but it is also a fundamental precondition for generating attosecond pulse trains and isolated attosecond pulses\,\cite{Krausz2009}. Therefore all future approaches towards attosecond diffractive imaging will require a multicolor analysis. We developed a multidimensional Simplex optimization\,\cite{Lagarias1998} on the basis of multicolor Mie scattering calculations\,\cite{Mie1908, Bohren1983} to analyze the diffraction images of spherical helium droplets and to demonstrate that the fits can give access to the optical properties of the droplets.

The majority of bright diffraction patterns ($\approx76\%$) shows ring structures that can be assigned to spherical droplets (cf Fig.\,3a). The diffracted field from dielectric spheres illuminated with a single wavelength can be described by the Mie solution and yields scattering patterns showing concentric rings\,\cite{Mie1908, Bohren1983}. The ring separation scales with the wavelength and with the inverse particle size. The overall scattering strength and the detailed shape of the ring structure further depend on the material's complex refractive index (see Methods). In our case, the XUV pulses contain spectral contributions from four harmonics (Fig.\,3d). Therefore the scattering pattern is a superposition of the corresponding single-wavelength scattering intensities, and displays a characteristic beating pattern due to the wavelength-dependent ring spacings of the individual spectral contributions to the image (black curve in Figs.\,3b/c). The observed patterns are fitted via a multidimensional Mie-based optimization with (i) the particle size, (ii) the refractive indices at the wavelengths of the contributing harmonics, (iii) the relative intensities of the harmonics and (iv) the intensity of the XUV pulse as input parameters (see Methods). While the optical properties of bulk liquid helium have been measured and calculated close to the 1s-2p transition of helium\,\cite{Surko1969, Lucas1983} (see Fig.\,3e), the dielectric function of the nanodroplets is completely unknown and expected to vary substantially with droplet size\,\cite{Joppien1993}. In fact, we find that fits using the bulk literature values for the refractive indices at the corresponding harmonic wavelengths cannot reproduce the observed diffraction patterns (see supplemental section 1\,\cite{Supplement}).

Successful fits can be achieved by using the refractive indices at the wavelengths of the dominant 13$^{\text{th}}$ and 15$^{\text{th}}$ harmonics as optimization parameters in addition to the particle size. In this procedure, the relative intensities of the contributing harmonics are set to measured average values and the refractive indices at the wavelengths of the 11$^{\text{th}}$ and 17$^{\text{th}}$ harmonics are fixed, as they lie far away from the large helium resonances\,\cite{Lucas1983}. The optimization was successfully carried out for 18 very bright scattering patterns with clear beating structures up to large scattering angles. The majority of these fits indicate a dominant contribution of the 13$^{\text{th}}$ harmonic, as exemplified in Fig.\,3b. However, for all patterns a second solution with dominant signal from the 15$^{\text{th}}$ harmonic is found by the algorithm with comparably low residuals of the  fit (cf. Fig.\,3c and supplemental section 1\,\cite{Supplement}). As the similarly intense 13$^{\text{th}}$ and 15$^{\text{th}}$ harmonics lie close to each other (only 3.1\,eV energetic distance), the fitting algorithm can vary the refractive indices freely and “exchange their roles” in the fit, while adjusting the cluster size accordingly (compare Figs.\,3b and c). In order to resolve such ambiguity in the optimization, future systematic studies are required with only one of the strong harmonics being near-resonance and/or with substantially better signal to noise ratio. By using higher energy and/or lower wavelength lasers to drive the HHG process, it is anticipated that the photon flux of individual harmonics can be further increased by at least one order of magnitude\,\cite{Popmintchev2015} while the energetic distance between the harmonics becomes larger. However, the present analysis supports that the multicolor fit procedure can be used for a new metrology of optical parameters and constitutes a basis towards future multicolor imaging approaches. In the subsequent scattering simulations, the average values from the solutions with dominant 13$^{\text{th}}$ harmonic are used (Fig.\,3b, cf. also supplemental information\,\cite{Supplement}).

Besides concentric ring patterns (Fig. 2a and b) and an abundance of about $20\%$ of elliptical patterns from ellipsoidal droplets (cf. Fig.\,2c), about $3\%$ of all bright images exhibit pronounced streak structures as exemplified in Fig.\,2 d-f and Fig.\,4a. The abundances of the three main types of patterns, i.e. rings, elliptical and streak patterns, are similar to what has been reported in previous hard X-ray measurements at LCLS\,\cite{Gomez2014}. However, whereas straight streak patterns were observed in the X-ray results, we find in the majority of our wide-angle scattering patterns a pronounced crescent-shaped bending of the streaks (statistics see Fig.\,2 and supplemental section 3\,\cite{Supplement}). In the X-ray experiments, the reconstruction of the corresponding droplet shapes via iterative phase retrieval revealed 2D projections of the helium droplets with extreme aspect ratios\,\cite{Gomez2014}. These were assigned to extremely flattened, "wheel-like" oblate shapes. The deformation was attributed to a high angular momentum, which can be transferred to the droplets by cavitation and rip-off from the liquid phase during the formation process\,\cite{Toennies2004, Gomez2014}, while vibrational excitations are assumed to decay very quickly\,\cite{SuppMatGomez2014}. Whereas classical viscid rotating droplets undergo a deformation from oblate to prolate two-lobed shapes with the rotation axis perpendicular to the long axis of the droplet\,\cite{Brown1980, Baldwin2015}, this transition has been suggested to be hindered in helium nanodroplets by the appearance of vortex arrays that deform the droplets and stabilize extreme oblate shapes\,\cite{Ancilotto2015}. Very recently the occurence of such classically unstable oblate helium droplets was further supported by statistical arguments, while an indication of rare prolate structures was also found\,\cite{Bernando2017}. However, a unique discrimination of prolate and oblate shapes based on the 2D projections accessible with small-angle X-ray diffraction is difficult\,\cite{Bernando2017}. Tomographic information, on the other hand, is contained in XUV wide-angle scattering and can be exploited to retrieve the three-dimensional particle shape and orientation provided the particle morphology is sufficiently regular\,\cite{Barke2015, Xu2014}. In our measurement, the presence of tomographic information prominently manifests in the bending of the streak patterns.

In order to retrieve the shapes underlying the experimentally observed scattering patterns, three dimensional multicolor scattering simulations were performed\,\cite{Sander2015} (see Methods). The two-sided, bent streak features can only be reproduced considering prolate droplets as shown in Fig.\,4b, matching the experimental pattern of Fig.\,4a. Crescent-shaped streaks arise from prolate structures that are tilted out of the scattering plane (i.e. the plane normal to the laser propagation axis). The wide-angle interference pattern can be intuitively understood in analogy to the reflection of a laser from a macroscopic rod. As shown in Fig.\,4c, bundles of rays diffracted by the cylindrical part of the surface gather the same path length and interfere constructively. In contrast, a tilted wheel-shaped particle with the same aspect ratio as exemplified in Fig.\,4d cannot explain the bending of the streaks. Instead, in the wide-angle scattering regime, the simulations indicate that such particles would generate a one-sided straight streak, a phenomenon that was not observed in our experiment. Although less obvious, also the analysis of observed straight streak patterns indicates that they can only be explained by prolate structures, as the observed streak signal is visible until the edge of the detector (cf. for example Fig.\,2e). Figs.\,4e,f show a comparison of diffraction patterns for pill and wheel shaped structures, that are aligned to the scattering plane. The tomographic nature of wide angle scattering reveals that the streaks decay much faster towards larger scattering angles for a wheel than for a pill-shaped particle. We would like to note that the absence of wheel-shaped droplets in our experiment may stem from the different droplet generation scheme, using a pulsed valve with a long nozzle in this work, where the transition to the superfluid state might be delayed compared to the short cw flow nozzle used by Gomez et al.\,\cite{Gomez2014}. However, our unambigous observation of prolate droplets, which are known to ocurr for classical liquids\,\cite{Brown1980, Baldwin2015}, will contribute to the discussion on the stability of spinning superfluid droplets. Their existence may provide a fascinating case for future experiments, as it should be possible to clarify if a prolate droplet shows macroscopic shape rotation, which is not expected for a superfluid droplet\,\cite{Ancilotto2015}.

We have shown the feasibility of single-shot single-particle coherent diffractive imaging using intense XUV pulses from a high harmonic generation source. Bright diffraction patterns of spherical helium droplets have been obtained and matched with simulations using optimized refractive indices. The observed crescent-shaped streak patterns could be uniquely assigned to prolate droplets. The results further suggest several future prospects connected to the HHG-specific properties. Laser-based HHG facilities provide a high accessibility compared to free-electron lasers in a wavelength regime suitable for 3D shape characterization of non-reproducible gas-phase nano-objects, particularly if experiments with single harmonics can be realized, which is anticipated using UV or deep UV driver lasers for HHG\,\cite{Cirmi2012,Popmintchev2015}. This will facilitate fundamental investigations of structure formation, such as tracing ice nucleation\,\cite{Levin2009}, with important implications for atmospheric physics and aerosol science. Moreover, unprecedented experiments beyond structural determination are possible, such as multicolor tomography and resonant-pump - resonant-probe CDI, that exploit the time-resolution and phase control achievable in HHG-based experiments for diffractive imaging of quantum coherent dynamics. The spatiotemporal characterization of ultrafast electron dynamics has been driving attosecond science\,\cite{Niikura2007,Krausz2009} from the beginning and will perhaps be the most exciting prospect of HHG-CDI. Considering the advancing capability of generating intense isolated attosecond pulses\,\cite{Takahashi2013} and the possibillity of stroboscopic illumination using attosecond pulse trains\,\cite{Mauritsson2008}, the vision of diffractive imaging of attosecond electron dynamics in isolated nanostructures has come in reach.

\begin{methods}

\subsection{Femtosecond laser system and generation of XUV harmonics.} The experiments are performed using a commercially available cryo-cooled Ti:sapphire laser amplifier (KMLabs Red Wyvern) delivering pulses at a central wavelength of 792\,nm with 33\,mJ pulse energy and 35\,fs pulse duration at a 1 \,kHz repetition rate. A fraction of 30$\%$ of the output energy (typically 10-12\,mJ) are taken for the generation of high harmonics. To this aim, a broadband spherical mirror with a focal distance of 5\,m is used to focus the near-infrared pulses into a 100\,mm long aluminum gas cell statically filled with $\approx$1.3\,mbar of xenon (loose focusing geometry). The position of the gas cell, the gas pressure and the NIR pulse energy are adjusted to optimize the HHG flux. An output energy of approximately 2\,$\mu$J (measured with a calibrated photodiode) is achieved in this geometry, corresponding to a conversion efficiency of 1.6$\times10^{-4}$ and an average power of 2\,mW. To the best of our knowledge, this is the highest average power obtained by means of HHG. The harmonic beam consists of the 11$^{\text{th}}$ (72\,nm), 13$^{\text{th}}$ (61\,nm), 15$^{\text{th}}$ (53\,nm), and 17$^{\text{th}}$ (47\,nm) harmonics (see Fig.\,3d) as measured by dispersing the XUV beam with a grating spectrometer. The XUV pulse duration was characterized in previous experiments using THz electron streaking technique to be roughly 20\,fs.

\subsection{Microfocusing setup and IR filter.} A high throughput XUV beamline (transmission $\approx10\%$)  consisting of three gold-coated, grazing incidence (10$^{\circ}$) toroidal mirrors and a flat Mo/Si mirror is used to tightly focus the XUV beam onto the target sample. Positions, radii and distances between the toroidal mirrors were optimized by ray tracing in order to achieve a high demagnification factor of 25 for the XUV beam while keeping the coma-aberrations low\,\cite{Frassetto2014,Poletto2013}. The first 40\,mm $\times10$\,mm toroidal mirror with radii 57.6\,m $\times$1.735\,m is placed 5\,m away from the gas cell in order to collimate the XUV beam. The collimated beam is then reflected by a flat Mo/Si mirror that partially absorbs the co-propagating near-infrared (NIR) laser pulse used for HHG. The remaining NIR laser beam ($\approx1$\,mJ) is filtered out by a 100\,nm thin aluminium filter. We note that the reduction of the NIR by the Mo/Si mirror is required to avoid damaging the aluminium foil. A coma-corrected system composed of two toroidal mirrors facing each other is then used to demagnify and tightly focus the XUV beam into the experimental chamber. The first of the latter two toroidal mirrors (radii 2650\,mm $\times$79\,mm) has a focal length of 230\,mm. Subsequently the last toroidal mirror (radii 3620\,mm $\times$109.2\,mm) is placed 680\,mm away from the focus to relay image the focus of the first toroidal mirror into the experimental chamber at a distance of 585\,mm. This geometry allows a demagnification factor of 25 with respect to the initial XUV spot size at the generation point. Considering a 0.5\,mrad $\times$ 0.5\,mrad diverging XUV beam with a 175\,$\mu$m  spotsize (FWHM) at the source point, we expect to achieve a minimum spot size at the focus of 7\,$\mu$m (FWHM). In our experiment, the size of the XUV beam was characterized by monitoring the fluorescence of a Cs/YAG screen placed at the focus using a CCD camera. We measured a 9\,$\mu$m$\times$10\,$\mu$m spot size (FWHM) with a 10\,$\mu$m beam waist, in close agreement with the expected value leading to an intensity of I$_f$ = 3$\times10^{12}$ W/cm$^2$.

\subsection{Helium droplet generation.} The helium nanodroplets are generated with a pulsed Even-Lavie valve\,\cite{Even2015} that is cooled with a Sumitomo closed-cycle cryostat down to 4.9\,K - 5.7\,K. The minimum temperature depends on the repetition rate and the opening duration of the valve (varied between 3 and 10\,Hz and 18 to 27\,$\mu$s, respectively) which influence the heat load of the valve. High-purity $^4$He (99.9999$\%$) at a pressure of 80 bars is expanded into a differentially pumped UHV chamber through a 100\,$\mu$m trumpet shaped nozzle located at 450\,mm distance to the interaction region. The droplet pulse is guided into the interaction chamber through a conical skimmer with 1 mm diameter, which reduces the uncondensed gas in the interaction chamber.

\subsection{Scattering experiment.} A large-area scattering detector ($\diameter$ 75\,mm) with a center hole ($\diameter$ 3\,mm) is placed 37\,mm behind the XUV focus, corresponding to a maximum spatial frequency of 0.09 nm$^{-1}$ for the dominant wavelength of 53\,nm. The detector consists of a Chevron-type MCP for signal amplification and a phosphor screen for conversion to optical light\,\cite{ClustersAtFLASH}. The MCP is used in pulsed operation to suppress background signal from charged particles. Further, the CDI-compatible focusing geometry described above allows for the use of two straylight apertures before the focus to minimize photonic background signal on the detector. The 8$^\circ$ tilt of the MCP channels results in an area with decreased response observable at the lower right side of the detector hole (cf. Fig.\,2, Fig.\,3a, Fig.\,4a)\,\cite{Fukuzawa2016}. The scattering patterns are recorded on a shot-to-shot basis using an out-of-vacuum CMOS camera. An ion time-of-flight spectrometer is used for establishing and optimizing the spatial overlap of the XUV pulses and the helium nanodroplets and the timing of the droplet jet\,\cite{Rupp2014}. Within $3\times10^5$ single-shot measurements 2300 bright patterns with distinct structures were obtained. Further 12700 recorded images contained weak, unstructured scattering signal. These statistics indicate that the experiment is performed in the single-particle limit as the probability to have two droplets in the focus at the same time is lower than $2\permil$.

\subsection{Data analysis and scattering simulations.}
For comparison with theory, the measured diffraction patterns were transformed to the scattered intensity that would be recorded on a spherical detector. In addition, the measured data must be corrected for the nonlinear detection efficiency of the MCP\,\cite{Bostedt2012,Barke2015}. Previous work has shown that the saturation effect can be described by an exponential efficiency function\,\cite{Barke2015} such that the detected signal intensity, $I_{\rm det}$, is connected to the true experimental intensity, $I_{\rm exp}$, via $I_{\rm det}=I_{\rm exp}^{\alpha}$. The nonlinearity exponent $\alpha$ = 0.5 has been found by matching the angular decay of the envelope of scattering profiles from spherical droplets to the universal $q^{-4}$ decay behavior predicted by Porod’s law\,\cite{Porod1951,Sorensen2000}. The center position was independently determined for every pattern to correct for slight variations resulting from wavefront tilts at the position of the droplet\,\cite{Loh2013}. Radial intensity profiles were extracted by angular averaging over the upper half of the detector to avoid any influence from the area on the MCP where the detection efficiency is decreased since the incoming photons impinge on the MCP parallel to the MCP channels\,\cite{Fukuzawa2016} (cf. Methods, Scattering experiment). In order to fit the measured patterns for spherical particles, we employed a multidimensional Simplex optimization\,\cite{Lagarias1998} on the basis of multicolor Mie scattering calculations\,\cite{Mie1908, Bohren1983}. The intensity pattern in a calculated multicolor image contains four single-frequency-components (11$^{\text{th}}$ to 17$^{\text{th}}$ harmonic), each weighted with the intensity of the respective harmonic order. As fitting parameters, the particle size and the photon energy-dependent refractive indices ($n=1-\delta-i\cdot\beta$ with $\delta$ being the deviation of the real part of the refractive index from unity and $\beta$, the imaginary part, which corresponds to the absorption) at the 13$^{\text{th}}$ and 15$^{\text{th}}$ harmonic were varied. For helium, the 11$^{\text{th}}$ and 17$^{\text{th}}$ harmonics are far away from resonances\,\cite{Surko1969, Lucas1983}, so that literature values of the refractive indices could be fixed within the fit procedure (17.2\,eV: $n = 0.97+i0.0$; 26.6\,eV: $n= 0.9964+i0.041$)\,\cite{NIST}. 18 measured patterns with clear rings up to maximum scattering angle were fitted using a hybrid Monte-Carlo Simplex optimization algorithm for a large ensemble of trajectories (see supplemental Figs.\,S5 and S6\,\cite{Supplement}). Each fitting trajectory was initialized with random start parameters in a reasonable range for the corresponding optimization parameters ($R=300$\,nm to $R=600$\,nm, $\delta=-0.3$ to $\delta=0.2$, $\beta=0$ to $\beta=0.07$), and subsequently improved via Simplex optimization. The scattering patterns for nonspherical shapes were calculated in the discrete-dipole-approximation as implemented in Ref.\,\cite{Sander2015} by the superposition of four single-color calculations and using the average optical parameters of the 13$^{\text{th}}$ and 15$^{\text{th}}$ harmonic determined in our study (solution with dominant 13$^{\text{th}}$ harmonic, 20.4\,eV: $n =
0.9252+i0.0178$; 23.5\,eV: $n = 1.2688+i0.0417$).

\end{methods}

\bibliographystyle{naturemag}

\begin{addendum}
 \item [Acknowledgements] The authors kindly acknowledge Bernd Sch{\"u}tte's excellent work for the initial development of the HHG beamline. The first author thanks Andrey Vilesov, Christoph Bostedt and Joachim Ullrich for helpful and enlightening discussions. Excellent support has been provided by the TUB-IOAP workshop. This project has received funding from DFG (Grants No. MO 719/13-1 and /14-1), from BMBF (Grant No. 05K13KT2), and from the European Union’s Horizon 2020 research and innovation programme under the Marie Sklodowska-Curie grant agreement No 641789. Further, T.F. acknowledges computational resources provided by the North-German Supercomputing Alliance (HLRN) and financial support from the Deutsche Forschungsgemeinschaft via SFB652/2, a Heisenberg Fellowship (ID: FE 1120/4-1) and from BMBF (grant ID: 05K16HRB). F.C. and M.N. acknowledge funding from ERC grants STARLIGHT (637756), and ELYCHE (227355).
 \item[Competing Interests] The authors declare that they have no competing financial interests.
 \item[Author Contributions] D.R. and Y.O. performed the feasibility studies in advance of the experiment. F.F., L.P., A.T., F.C. and M.N. developed the microfocusing optics setup and implemented it together with N.M. and A.R.  B.L. set up the helium jet, D.R. and M.S. set up the CDI detection system, and J.Z. developed the data acquisition system. N.M. and A.R. operated the HHG source, and D.R., M.S., B.L., N.M. and A.R. assembled and carried out the experiment. K.S., C.P., and T.F. developed and performed the scattering simulations. D.R., N.M., B.L., J.Z., K.S., C.P., and T.F. analyzed the data with input from all authors. The manuscript was discussed and written with input from all authors.
 \item[Correspondence] Correspondence and requests for materials should be addressed to D.R., T.F., and A.R.~(email: daniela.rupp@physik.tu-berlin.de, thomas.fennel@uni-rostock.de, arnaud.rouzee@mbi-berlin.de).
\end{addendum}
\clearpage

\begin{figure}
\begin{center}
\includegraphics[width=0.9\textwidth,keepaspectratio=true]{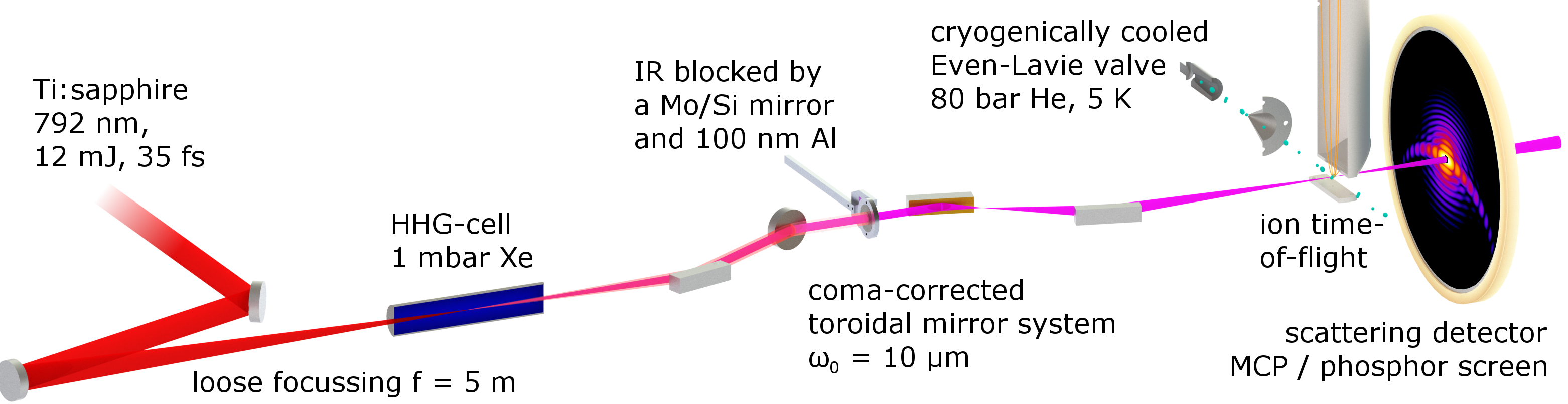}
\end{center}
\caption{\textbf{Scheme of the experimental setup}. A Ti:sapphire laser with 792\,nm central wavelength and 35\,fs pulse duration is used for the generation of high harmonics. Up to 12\,mJ are loosely focused into a xenon-filled cell, where the XUV pulses are produced. After removing the copropagating NIR with a Mo/Si mirror and an aluminum filter, the beam is focused to a small spot ($\omega_0=10$\,$\mu$m) using a coma-correcting system of gold-coated toroidal mirrors\,\cite{Frassetto2014}. A pulsed jet of helium nanodroplets ($\overline{R}\approx$ 400\,nm) is overlapped with the XUV focus. The overlap is optimized by monitoring the formation of He$^+$ ions using an ion time-of-flight spectrometer. The scattering signal is amplified by a pulsed MCP and converted to optical photons on a phosphor screen. The single-shot diffraction images are captured with an out-of-vacuum camera (not depicted).}
\end{figure}

\begin{figure}
\begin{center}
\includegraphics[width=0.5\textwidth,keepaspectratio=true]{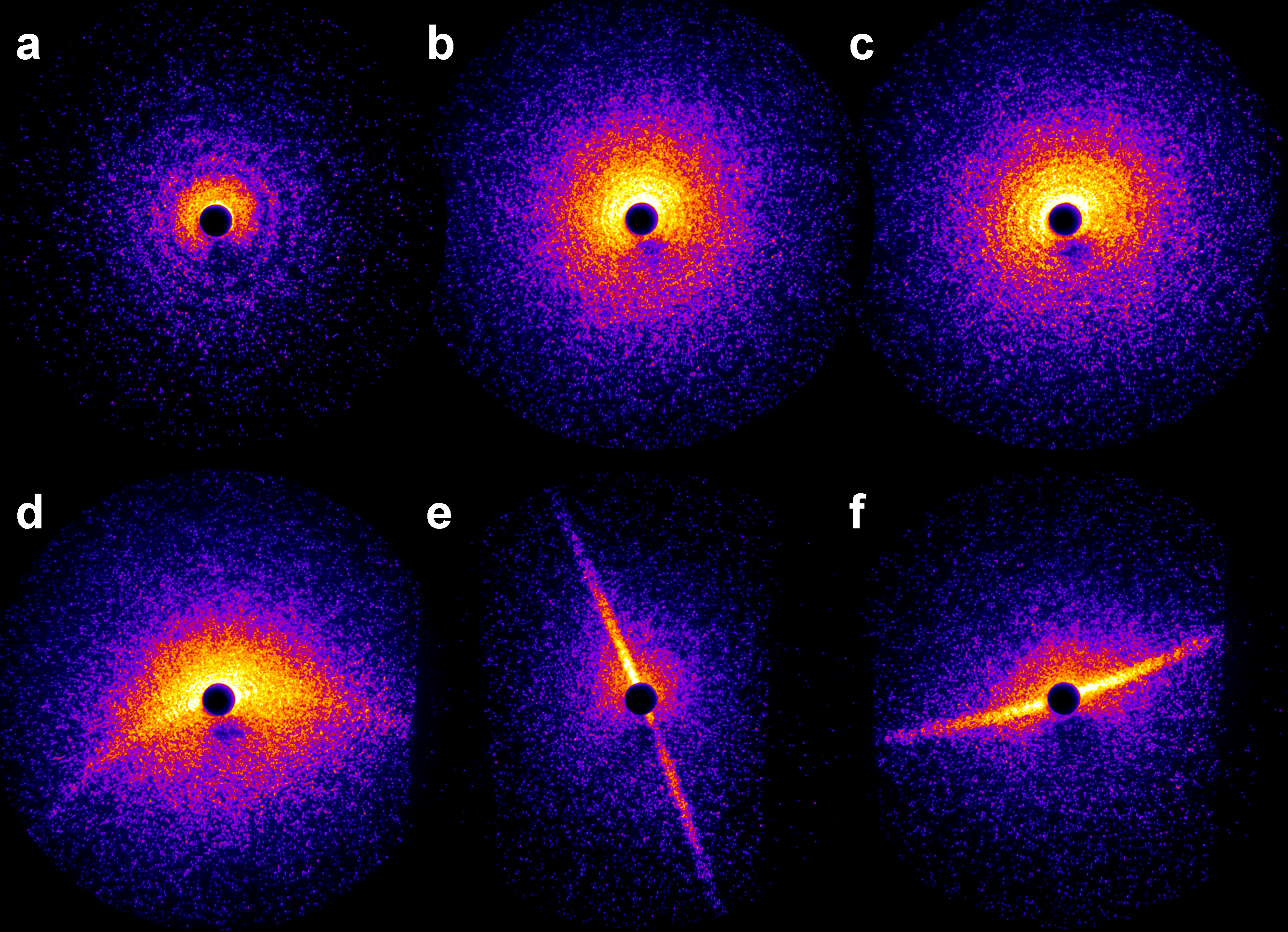}
\caption{\textbf{Three characteristic types of diffraction images from individual helium nanodroplets}. The majority of images contains concentric ring patterns (a,b) that are assigned to spherical droplets. Elliptical ring structures (c) or pronounced streak patterns (d-f) reflect deformed helium droplets. The measured dataset comprises 1762 ring-type, 421 elliptical, and 68 streak-type images. In most cases, the latter exhibit a clear bending of the streaks (55 out of 68 images), e.g. as in (d) and (f). For details see the supplementary information\,\cite{Supplement}.}
\end{center}
\end{figure}

\begin{figure}
\begin{center}
\includegraphics[width=0.9\textwidth,keepaspectratio=true]{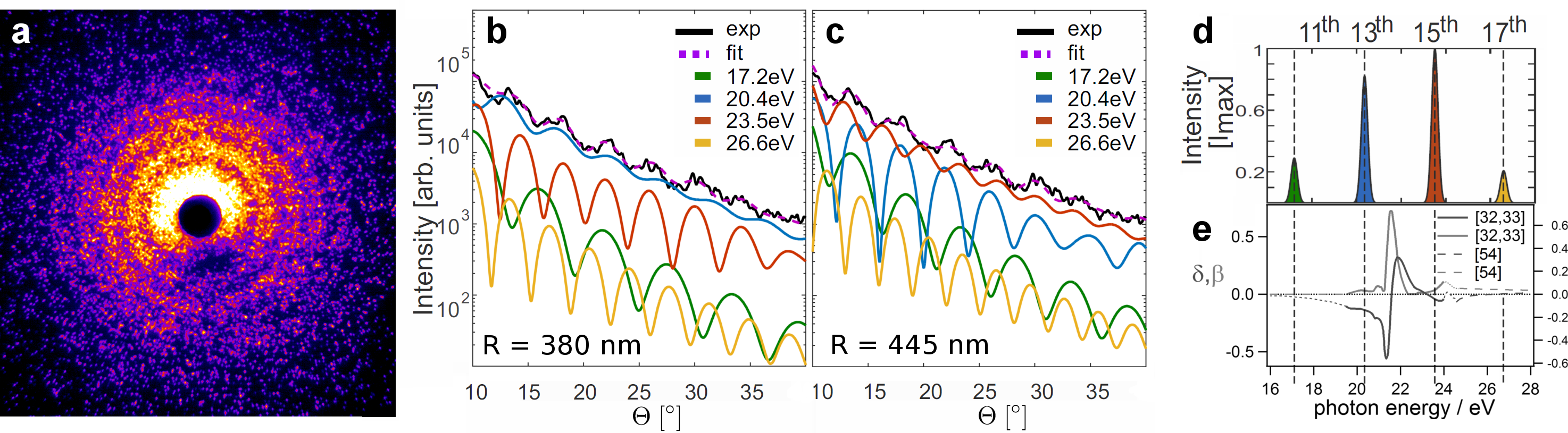}
\end{center}
\caption{\textbf{Multicolor analysis of the diffraction images}. (a) Measured bright scattering image from a spherical droplet with pronounced concentric ring pattern. (b,c) Multicolor Mie fits (dashed purple) of the measured radial intensity profile (solid black) from (a) as obtained via a simplex optimization (see Methods) of the individual harmonic contributions to the profiles (color-coded in green, blue, orange, and yellow). The results illustrate that two qualitatively different solutions yield comparably small residuals. The two solutions indicate that either the 13$^{\text{th}}$ harmonic (b) or the 15$^{\text{th}}$ harmonic (c) yields the dominant contribution (see text). (d) Measured average XUV spectrum of the HHG radiation. (e) Sketch of the energy dependent refractive indices of bulk liquid helium in the vicinity of the helium 1s-2p transition, assembled from literature values\,\cite{Surko1969, Lucas1983, NIST}.}
\end{figure}

\begin{figure}
\begin{center}
\includegraphics[width=0.65\textwidth,keepaspectratio=true]{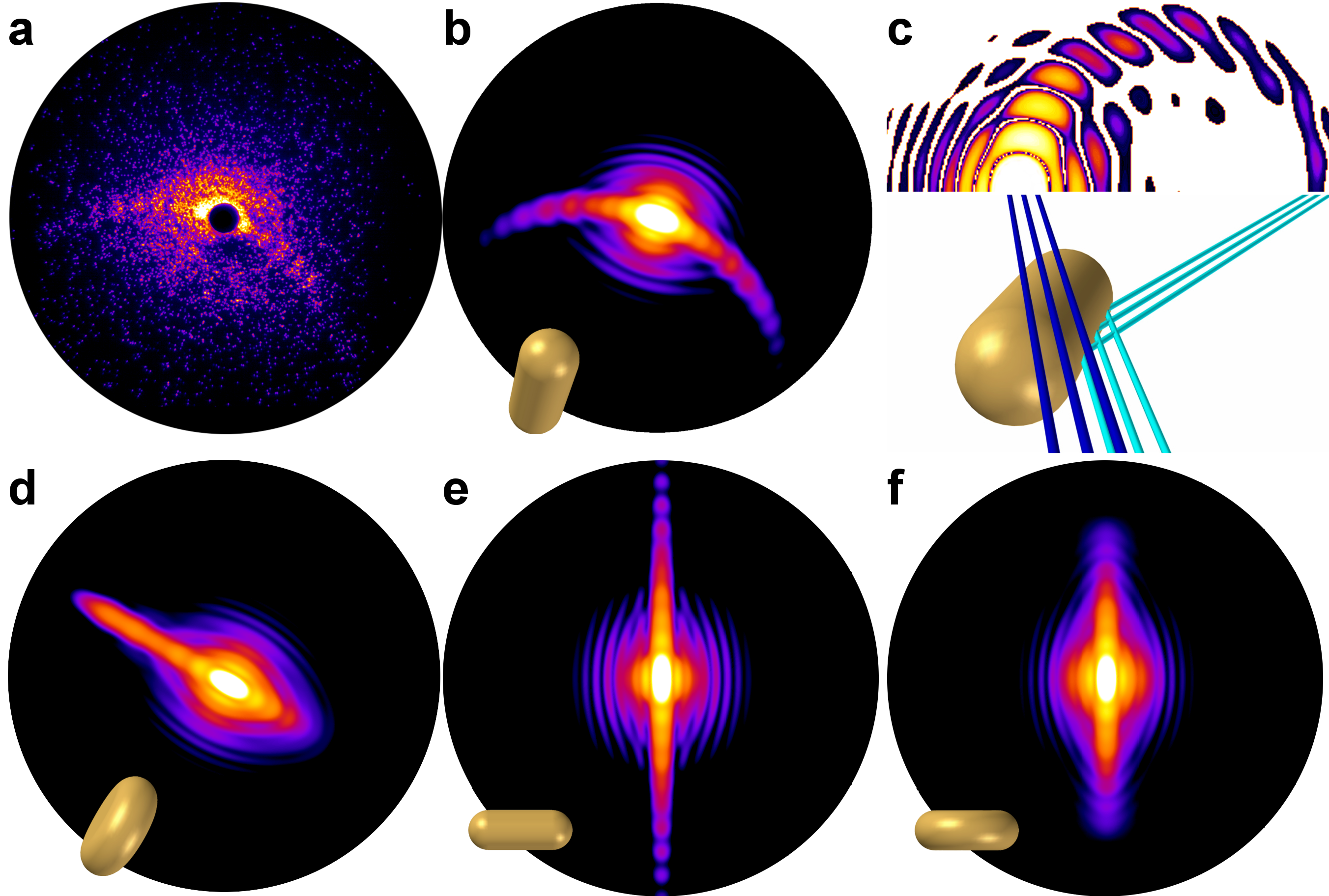}
\caption{\textbf{Unique identification of prolate, pill-shaped structures}. (a) Measured image and (b) matching simulation result of the wide-angle diffraction of a pill-shaped prolate droplet. Simulation parameters: 35$^\circ$ tilt angle with respect to the optical axis, two short half-axes $a=b=370$\,nm, one long half-axis $c= 950$\,nm. (c) Illustration of the origin of bent streaks occurring when a tilted rod-type structure diffracts the light. Two particular bundles of constructively interfering rays are explicitly sketched (the different ray colors were chosen for visibility, they do not refer to wavelenghts). (d) Simulated wide-angle diffraction image of a wheel-shaped oblate particle. If the particle’s symmetry axis is tilted out of the diffraction plane, the diffraction patterns exhibit straight streaks to only one side (long and short half-axes as in (b), tilt angle 10$^\circ$, further parameters see Methods). (e,f) Comparison of simulated wide-angle diffraction images of a prolate (e) and an oblate structure (f) aligned to the scattering plane. Though the 2D projections are similar and 2D outlines identical, the intensity distributions of the straight streaks are clearly different and decay much faster for "wheels" than for "pills".}
\end{center}
\end{figure}

\end{document}